\documentclass{sig-alt-release2}

\newcommand{\sys}{RDFViewS}

\begin{document}

% --- Author Metadata here ---
\conferenceinfo{CIKM'10,} {October 25--29, 2010, Toronto, Ontario, Canada.} 
\CopyrightYear{2010}
\crdata{978-1-4503-0099-5/10/10}
\clubpenalty=10000
\widowpenalty = 10000 
% --- End of Author Metadata ---

\title{\sys: A Storage Tuning Wizard for RDF Applications\titlenote{This work has been partially funded by Agence Nationale de la Recherche, decision ANR-08-DEFIS-004.}}

% \numberofauthors{4}
\author{Fran\c cois Goasdou\'e$^1$, Konstantinos Karanasos$^1$, Julien Leblay$^{1,2}$, and Ioana Manolescu$^1$ \vspace{3mm}\\
$^1$Leo team, INRIA Saclay and LRI, Universit\'e Paris-Sud 11, Orsay, France\\
$^2$Universit\'e Paris 6, Paris, France\\
{\email{firstname.lastname@inria.fr}}}

\date{}

\maketitle

\begin{abstract}
In recent years, the significant growth of RDF data used in numerous applications has made its efficient and scalable manipulation an important issue.
In this paper, we present \sys, a system capable of choosing the most suitable views to materialize, in order to minimize the query response time for a specific SPARQL query workload, while taking into account the view maintenance cost and storage space constraints. 
Our system employs practical algorithms and heuristics to navigate through the search space of potential view configurations, 
and exploits the possibly available semantic information - expressed via an RDF Schema - to ensure the completeness of the query evaluation.
\end{abstract}

\vspace{1mm}
\noindent
{\bf Categories and Subject Descriptors:} H.3.4 {Information Storage And Retrieval}: {Systems and Software}; H.2.1 {Database Management}: {Logical Design}

\vspace{1mm}
\noindent
{\bf General Terms:} Algorithms, Design, Performance

\vspace{1mm}

\section{Outline}
\label{sec:outline}

RDF data is increasingly used in data management applications related to traditional Computer Science topics (e.g., search engines, semantic annotations, social tagging), as well as in contexts well beyond this traditional scope (e.g., RDF is becoming prevalent in many Life Science and in particular BioInformatics applications). These and other applications have significantly increased the volumes of RDF data to be handled. This size effect and the complexity and irregularity of RDF data pose significant challenges to the task of building an efficient query evaluation engine.

RDF data consists of triples of the form (\emph{subject}, \emph{property}, \emph{object}). This seemingly simple data model leads to complex queries and expensive evaluation, since any meaningful question requires forming chains of several triples, 
which are translated to many-join queries over a single, huge table containing all the triples. One approach taken in order to handle such large data volumes consists of mapping the data into one or several relations, and storing them in a relational database management system (RDBMS), possibly endowed with specific indexes~\cite{1325900,DBLP:journals/pvldb/WeissKB08}. Then, RDF queries expressed in SPARQL can be translated to SQL queries~\cite{DBLP:journals/dke/ChebotkoLF09}, which are evaluated by the RDBMS. Another approach consists of developing RDF-specific stores and query processors~\cite{1559911}, which still share some of the standard notions and features of relational storage engines.

%---------------
\begin{figure}[t]
\begin{center}
% \vspace{-6mm}
\includegraphics[width=0.80\columnwidth]{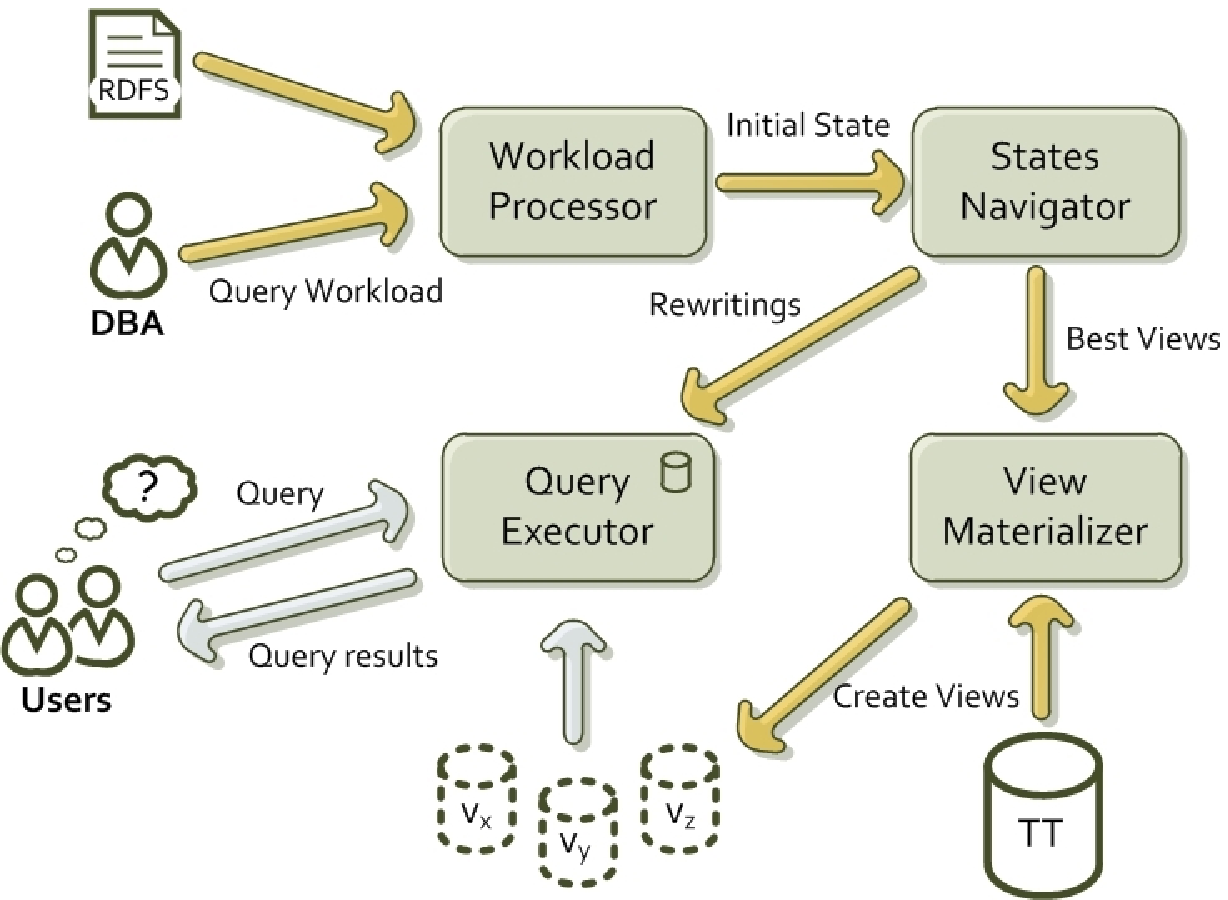}
\vspace{-2mm}
\caption{\sys\ architecture.\label{fig:system}}
\end{center}
\vspace{-9mm}
\end{figure}
%---------------

These efforts aim at providing a generic, one-size-fits-all storage model for RDF. However, decades of research and development of RDBMSs has shown that huge performance gains can be achieved by tuning the storage to the data sets and to the requirements of specific applications. This is typically achieved by establishing {\em materialized views and/or indices} specific to the data and workload~\cite{DBLP:journals/dke/TheodoratosLS01}. Another important aspect of RDF data management is that the rich semantic information, under the form, e.g., of an RDF Schema, can be associated to the data set. In this situation, schema-based reasoning may lead to finding answers to a query, which would simply not be found by querying the data alone. Thus, the interpretation of RDF queries may be affected by the existence of associated semantics, and this must be taken into account when designing a query-inspired set of views.

We propose to demonstrate \sys\ (standing for \emph{RDF} \emph{View} \emph{S}election), a system that focuses on automatically choosing the materialized views which are most appropriate for a given data set and query workload. The tool provides many options to guide the search, into which it also incorporates the insights brought by an RDF Schema, if one is available. \sys\ outputs a set of proposed materialized views (which are automatically created within an RDBMS), as well as a set of rewritings (or reformulations) of the original workload, in terms of these materialized views. Thus, \sys\ can, in effect, be seen as a storage tuning wizard for RDF data, to be used in conjunction with off-the-shelf RDBMSs. The tool's various steps and options can be easily inspected and controlled via a GUI by \sys ' target users: administrators of large RDF databases. 

\vspace{-3mm}

\section{Problem Model}
\label{sec:model}

\sys\ takes as input a set of conjunctive SPARQL queries. Each query is endowed with a weight, reflecting its relative importance (e.g. how often it is posed).

We model our problem as a search state optimization problem, based on an existing proposal for selecting views to materialize in a relational setting~\cite{DBLP:journals/dke/TheodoratosLS01}, which we adapted to the particularities of the RDF model. For a given query workload $Q$, we define a state as the pair $S_i(Q) = \langle V_i,R_i\rangle$, where $V_i$ is the set of views to materialize and $R_i$ the rewritings needed to answer the queries of $Q$ using exclusively the views in $V_i$. 

We use three transitions, which can be applied to a given state and yield a new one: {\em selection cut}, {\em join cut} and {\em view fusion}. Intuitively, the first two aim at relaxing the queries, by removing some predicates. The third one attempts to fuse two candidate views, replacing them by a single one. The relaxation steps help eliminate constraints which differentiate two views, so that view fusion may be applied. If the workload queries have common sub-queries, these will be identified as useful views to materialize.

The quality of each state is assessed using a {\em quality function}, which reflects the query execution time, the view maintenance cost and the space needed for materializing the views of the state.
Starting from an initial state, we apply the transitions and navigate in the search space according to a search strategy. As the \emph{initial state} of our search, we choose the one that proposes to materialize exactly the query workload (best execution time, worst view maintenance cost). At the end of the search we return the state with the best quality score (minimum combined cost).

\vspace{-3mm}

\section{System Architecture}
\label{sec:system}

The architecture of \sys\ is depicted in Figure~\ref{fig:system}. The RDF data is initially stored into an RDBMS as a single triple table ({\em TT}); for efficiency, and following many similar works~\cite{1453927}, the table is dictionary-encoded, i.e., URIs and string constants are assigned distinct integers, and the {\em TT} table stores triples of integers. The database administrator ({\em DBA}) uses \sys\ to further tune the store. To this end, she provides the SPARQL query workload to the \textbf{Workload Processor} through a graphical interface. In the presence of an RDF Schema, the queries are reformulated, compiling the knowledge of the Schema inside them and transforming each query to a union of queries~\cite{RDFVS-TR}. The (possibly reformulated) queries are used to create the \emph{initial state} of the search.

The initial state is then loaded to the \textbf{States Navigator}, which constitutes the gist of our system. We have devised two exhaustive strategies that navigate through the whole search space. However, as the problem we address is known to be well above exponential, we employ heuristics which significantly prune the search space. Moreover, we provide the option to apply some additional stop conditions: we identify states that have some specific characteristics and we do not allow more transitions to be applied on these states. More details about the search strategies can be found in~\cite{RDFVS-TR}.

Once the search is finished, we obtain the best state according to our quality function and we materialize the views of this state, after translating them to SQL (\textbf{View Materializer}). Then, we push the rewritings contained in the best state to the \textbf{Query Executor}, which stores them for future use. Whenever a user issues a query from the workload, the Query Executor uses the stored rewritings to efficiently answer the query by using the already materialized views.

\vspace{-3mm}

\section{Demonstration Scenario}
\label{sec:scenario}

Our system has been fully implemented in Java 6. The triple table and the materialized views are stored in PostgreSQL v8.4.4.  We have built a web-based interface which enables users to interact with the system, extensively parameterize it and follow in detail the view selection process.  Screen captures and further details on the system can be found at the \sys\ website\footnote{http://rdfvs.saclay.inria.fr}.

Demo attendees will play the role of a database administrator. Using the interface, they will first choose one of the pre-loaded RDF datasets (among others, some of the the most widely-used RDF datasets will be available: Barton, Yago, Uniprot and LUBM), and the query workload for which they want to tune the database. They may also load their own datasets, modify the existing query workloads or add new ones. The queries can be modified either by using a SPARQL editor or through a visual editor we have created. Finally, they will pick the RDF Schema(s) they wish to use.

Before initializing the search for the best view configuration, attendees will define some additional details of the searching process, according to their specific preferences. In particular, they will choose whether they prefer a quick search, or a search that lasts longer but guarantees the optimal solution. Furthermore, they will tune the quality function used by adjusting the weights of its components (giving more importance to the query execution time, to the view maintenance or to the space needed).

After the end of the search, the selected views are displayed, together with their space cost and performance gains. Moreover, a graphical overview of the search space will be given. This information will also act as a feedback to the user, which may choose to tune differently the quality functions in a subsequent search etc. 

To verify the performance benefits brought by \sys, attendees will then act as simple users issuing queries, which will be first answered against the triple table and then by exploiting the materialized views.


\begin{thebibliography}{1}

\vspace*{0.5mm}
\scriptsize
 
\bibitem{1325900}
D.~J. Abadi, A.~Marcus, S.~R. Madden, and K.~Hollenbach.
\newblock Scalable semantic web data management using vertical partitioning.
\newblock In {\em VLDB}, 2007.

\bibitem{DBLP:journals/dke/ChebotkoLF09}
A.~Chebotko, S.~Lu, and F.~Fotouhi.
\newblock Semantics preserving {SPARQL-to-SQL} translation.
\newblock {\em Data Knowl. Eng.}, 68(10), 2009.

\bibitem{RDFVS-TR}
F.~Goasdou\'e, K.~Karanasos, J.~Leblay, and I.~Manolescu.
\newblock Materialized view-based processing of {RDF} queries.
\newblock http://rdfvs.saclay.inria.fr/RDFViewS-TR.pdf, 2010.
\newblock {Technical report}.

\bibitem{1453927}
T.~Neumann and G.~Weikum.
\newblock {RDF-3X}: a {RISC}-style engine for {RDF}.
\newblock {\em Proc. of VLDB}, 1(1), 2008.

\bibitem{1559911}
T.~Neumann and G.~Weikum.
\newblock Scalable join processing on very large {RDF} graphs.
\newblock In {\em SIGMOD}, New York, NY, USA, 2009. ACM.

\bibitem{DBLP:journals/dke/TheodoratosLS01}
D.~Theodoratos, S.~Ligoudistianos, and T.~K. Sellis.
\newblock View selection for designing the global data warehouse.
\newblock {\em Data Knowl. Eng.}, 39(3), 2001.

\bibitem{DBLP:journals/pvldb/WeissKB08}
C.~Weiss, P.~Karras, and A.~Bernstein.
\newblock Hexastore: sextuple indexing for semantic web data management.
\newblock {\em Proc. of VLDB}, 1(1), 2008.

\end{thebibliography}
\end{document}